\documentstyle[preprint,aps,epsf]{revtex}
\tighten

\begin{document}

\draft

\title{Polarization observables in the semiexclusive photoinduced three-body breakup of $\bbox{^3}$He.}

\author{
R.~Skibi\'nski$^1$,
J.~Golak$^{1}$,
H.~Wita\l{}a$^1$,
W.~Gl\"ockle$^2$,
A.~Nogga$^3$,
H.~Kamada$^4$
}
\address{$^1$M. Smoluchowski Institute of Physics, Jagiellonian University,
                    PL-30059 Krak\'ow, Poland}
\address{$^2$Institut f\"ur Theoretische Physik II,
         Ruhr-Universit\"at Bochum, D-44780 Bochum, Germany}
\address{$^3$ Forschungszentrum J\"ulich, Institut f\"ur Kernphysik (Theorie), D-52425 J\"ulich, Germany}
\address{$^4$ Department of Physics, Faculty of Engineering,
   Kyushu Institute of Technology,
   1-1 Sensuicho, Tobata, Kitakyushu 804-8550, Japan}

\date{\today}
\maketitle

\begin{abstract}
The photon and $^3$He analyzing powers as well as spin correlation coefficients in the semiexclusive three-body 
photodisintegration of $^3$He are investigated for incoming photon laboratory energies
E$_\gamma$=12, 40 and 120 MeV. 
The nuclear states are obtained by solving three-body Faddeev equations with the AV18 nucleon-nucleon
potential alone or supplemented with the UrbanaIX three-nucleon force. 
Explicit $\pi$- and $\rho$-meson exchange currents 
are taken into account, but we also compare to other models of the electromagnetic current. 
In some kinematical conditions we have found strong effects of the three-nucleon force for the $^3$He analyzing power and
spin correlation coefficients, as well strong sensitivities to the choice of the currents.
This set of predictions should be a useful guidance for the planning of measurements.
In addition, we compare our results for two-body $^3$He breakup induced by polarized photons
with a few existing data.

\end{abstract}
\pacs{21.45.+v, 24.70.+s, 25.10.+s, 25.20.-x}

\narrowtext

\section{Introduction}
\label{secI}

The study of polarization phenomena is a natural extension of investigation of unpolarized processes.
It provides additional information on details of the underlying nuclear Hamiltonian not
available in unpolarized reactions. 
In nucleon-nucleon (NN) systems the polarized processes provide a necessary 
data set to construct the NN potentials~\cite{PSA}.
The investigation of nucleon-deuteron (Nd) elastic scattering and the deuteron breakup reaction with 
polarized incoming nuclei or polarization of the outgoing particles measured 
is indispensable to learn about properties of the three-nucleon (3N) forces.
Nowadays spin observables in Nd elastic scattering where the initial deuteron and/or
nucleon is polarized and also the polarization of the final particles is measured, are available 
and can be compared with rigorous theoretical predictions\cite{WitalaNdelastic,Witalamagnetic,chiral,Kievsky}. 
Also for deuteron breakup such studies were performed, both experimentally~\cite{Allet,Qin,Stephan}
as well as theoretically~\cite{KurosI,KurosII}. 
Altogether, this allowed to test the current models
of the nuclear Hamiltonian.

In addition to the strong forces, the electromagnetic processes contain new dynamical 
ingredients due to the interaction between real or virtual photons with the currents of nuclei.
It was found that in such processes contributions to the nuclear current due to meson 
exchanges play an important role.
Studies of polarization observables in photo- and electro-induced processes on the deuteron~\cite{arenhovel},
as well as in the Nd radiative capture
\cite{golak2000,skibinski1,viviani,hannover,marcucci2005} can be used to determine the structure 
of nuclear currents. The combination of strong and electromagnetic interactions 
is a demanding test for theoretical models. 
The results up to now show an overall good agreement of theoretical predictions with the data,
however there is still room for improvement~\cite{raport2005}.
Recently an important progress is observed in experimental investigations of
processes with polarized photons. The high-intensity sources of highly polarized 
photon beams obtained by the Compton backscattering give hope for future 
precise data ~\cite{tunl}. The analysis of the first measurement of the $^3$He breakup using
polarized photons at low energies is in progress and was reported recently
in~\cite{groningentunl}.

In this paper we would like to present the results of theoretical  
investigations of spin observables in kinematically incomplete $\vec{\gamma} (^3\vec{He},N)NN$ processes
in which the incoming photon and/or the $^3$He nucleus are polarized. 
This study is done for three photon laboratory energies $E_\gamma$=12, 40 and 120 MeV.
For each photon energy, the energy spectrum of the detected outgoing nucleon 
at different angles has been calculated. 
We restrict ourselves to photon energies below the pion production threshold
and have chosen the above energies as examples of low, intermediate and
high photon energies. It was shown in ~\cite{skibinski2} that for those energies
one can expect different manifestations of the action of the 3N force in two-body
photodisintegration of $^3$He. While at low energies the inclusion of the three-nucleon forces 
decreases the cross section, at higher energies 3N forces act in the opposite direction.
At intermediate energies the influence of 3N forces on the two and three-body breakup 
cross sections is negligible. 
As will be shown in section~\ref{secIII} for several spin observables
the influence of 3N forces is visible in the semiexclusive spectrum of the outgoing nucleon 
also at intermediate energies of the incoming photon. The presented results 
should be a useful guide for 
future experiments. 
Up to now, to the best of our knowledge,  no such predictions have been published.
 
In section ~\ref{secII} we shortly describe the theoretical formalism underlying our calculations 
and give definitions for the studied spin observables. In section ~\ref{secIII} we present
our predictions for three-body breakup. In addition we turn into two-body $^3$He breakup and 
compare our results to a few existing data.
We summarize in section ~\ref{secIV}.

\section{Theoretical Framework}
\label{secII}
The theoretical framework we use is described in detail in Refs.~\cite{skibinski1,raport2005,skibinski2,report}.
For the convenience of the reader we briefly summarize the most important steps.
The basic nuclear matrix element $N_{m_i,\tau,m}^{\rm 3N}$ is expressed through the state  
$\mid \tilde{U}_{\tau}^{m} \rangle$ which fulfills the  Faddeev-like equation
\begin{eqnarray}
\mid \tilde{U}_{\tau}^{m} \rangle  \ = \
( 1 + P ) j_\tau (\vec Q ) \mid \Psi_{^3{\rm He}}^m \rangle \ + \
\left( t G_0 P + \frac12 ( 1 + P ) V_4^{(1)} G_0 ( t G_0 +1) P \right)
\mid \tilde{U}_{\tau}^{m} \rangle .
\label{eq:Utilde}
\end{eqnarray}
Here $ j_\tau (\vec Q ) $ is a spherical $\tau-$component of the $^3$He electromagnetic current operator,
$t$ the NN t-matrix,
$G_0$ the free 3N propagator and $P$ the sum of the cyclical and
anticyclical permutations of 3 particles. Further $V_4^{(1)}$
is that part of the 3NF, which is symmetrical (like the NN $t$-matrix) under the exchange
of nucleons 2 and 3, and $\mid \Psi_{^3{\rm He}}^m \rangle$ is the $^3$He bound state with spin projection $m$.
The nuclear matrix element for three-body breakup of $^3$He is given via 
\begin{eqnarray}
N_{m_i,\tau,m}^{\rm 3N} =
\frac12  \langle \Phi_0^{m_i} \mid ( t G_0 + 1 ) P \mid \tilde{U}_{\tau}^{m} \rangle ,
\label{eq:N3NN}
\end{eqnarray}
where $ \langle \Phi_0^{m_i} \mid $  is the properly
anti-symmetrized (in the two-body subsystem) state of three free nucleons with their spin projections $m_i$.

Given the $N_{m_i,\tau,m}^{\rm 3N}$ amplitudes, one can calculate any polarization observables.
They are expressed through the nuclear matrix elements with different 
spin projections carried by the initial photon, the $^3$He nucleus, and by
the outgoing nucleons.

Chossing the z-axis to be the direction of the incoming photon and allowing for a linear photon polarization
$P_0^{\gamma}$ along the x-axis,  
with the polarization component $P_0^{\gamma}=-1$,  and for the $^3$He target
nucleus polarization $P_0^{^3He}$ along the y-axis, 
the cross section in a kinematically incomplete reaction $\vec{\gamma} (^3\vec{He},N)NN$, 
when the outgoing nucleon is detected 
at angles ($\theta,\phi$) is given by
\begin{eqnarray}
\sigma^{pol}_{\gamma,^3He} (\theta, \phi) &=&  \sigma^{unpol}_{\gamma,^3He} (\theta) [ 1 +
P_0^{\gamma} ~ cos(2\phi) ~ A_x^{\gamma}(\theta) ~ + ~
 P_0^{^3He} ~ cos(\phi) ~ A_y^{^3He}(\theta) ~ + ~ \cr
&~&P_0^{\gamma}~ cos(2\phi) ~ P_0^{^3He} ~ cos(\phi) ~C_{x,y}^{\gamma,^3He}(\theta)
~ + ~
P_0^{\gamma} ~ sin(2\phi) ~ P_0^{^3He} ~ sin(\phi) ~C_{y,x}^{\gamma,^3He}(\theta)
 ].
\label{eq1}
\end{eqnarray}
Here the nonvanishing spin observables are the photon ($A_x^{\gamma}(\theta)$) 
and the $^3$He ($A_{y}^{^3He}(\theta)$) analyzing powers, 
and the spin correlation coefficients $C_{x,y}^{\gamma,^3He}(\theta)$ and $C_{y,x}^{\gamma,^3He}(\theta)$.
They can be obtained by measuring the spectra of the outgoing nucleon using a proper 
combination of $\phi$ angles and are expressed through the nuclear matrix element $N_{m_i,\tau,m}^{\rm 3N}$ by:
\begin{eqnarray}
A_x^{\gamma}(\theta)&\equiv& {{ \sum_{m_i m} (2\Re\lbrace N_{m_i, -1 m}
N_{m_i, +1 m}^*\rbrace) }
\over{\sum_{m_i m}
({\vert N_{m_i, +1 m} \vert }^2 + {\vert N_{m_i, -1 m} \vert }^2 ) }} \cr
A_y^{^3He}(\theta)&\equiv&
{ { \sum_{m_i} (-2\Im\lbrace N_{m_i, -1  -{1\over{2}} }
N_{m_i, -1 {1\over{2}} }^*\rbrace - 2\Im\lbrace N_{m_i, +1 -{1\over{2}} }
N_{m_i, +1 {1\over{2}} }^*\rbrace )   }
\over{\sum_{m_i m}
({\vert N_{m_i, +1 m} \vert }^2 + {\vert N_{m_i, -1 m} \vert }^2 ) }} \cr
C_{x,y}^{\gamma,^3He}(\theta)&\equiv&
{ { \sum_{m_i} (-2\Im\lbrace N_{m_i, -1 -{1\over{2}} }
N_{m_i, +1 {1\over{2}} }^*\rbrace + 2\Im\lbrace N_{m_i, -1 {1\over{2}} }
N_{m_i, +1 -{1\over{2}} }^*\rbrace )   }
\over{\sum_{m_i m}
({\vert N_{m_i, +1 m} \vert }^2 + {\vert N_{m_i, -1 m} \vert }^2 ) }} \cr
C_{y,x}^{\gamma,^3He}(\theta)&\equiv&
{ { \sum_{m_i} (2\Im\lbrace N_{m_i, -1 -{1\over{2}} }
N_{m_i, +1 {1\over{2}} }^*\rbrace + 2\Im\lbrace N_{m_i, -1 {1\over{2}} }
N_{m_i, +1 -{1\over{2}} }^*\rbrace )   }
\over{\sum_{m_i m}
({\vert N_{m_i, +1 m} \vert }^2 + {\vert N_{m_i, -1 m} \vert }^2 ) }} 
\label{eq2}\;.
\end{eqnarray}

\section{Results}
\label{secIII}

We solved Eq.(\ref{eq:Utilde}) using a momentum space partial wave decomposition and
the AV18 nucleon-nucleon  
potential~\cite{ref.AV18} 
alone or supplemented with the Urbana IX 3NF~\cite{ref.urbanaIX}. For both parities and the total
angular momentum 
of the 3N system 
$J \leq \frac{15}{2}$ all partial waves with angular momenta 
in the two-body subsystem 
$j \leq 3$ have been used. We refer to~\cite{report} for more details on our 
basis, partial wave decomposition and numerics.
The electromagnetic nuclear current operator was taken as the single nucleon current 
supplemented by the exchange currents of the $\pi$- and $\rho$-like nature~\cite{golak2000}.

\begin{figure}[h!]
\leftline{\mbox{
\epsfysize=190mm \epsffile{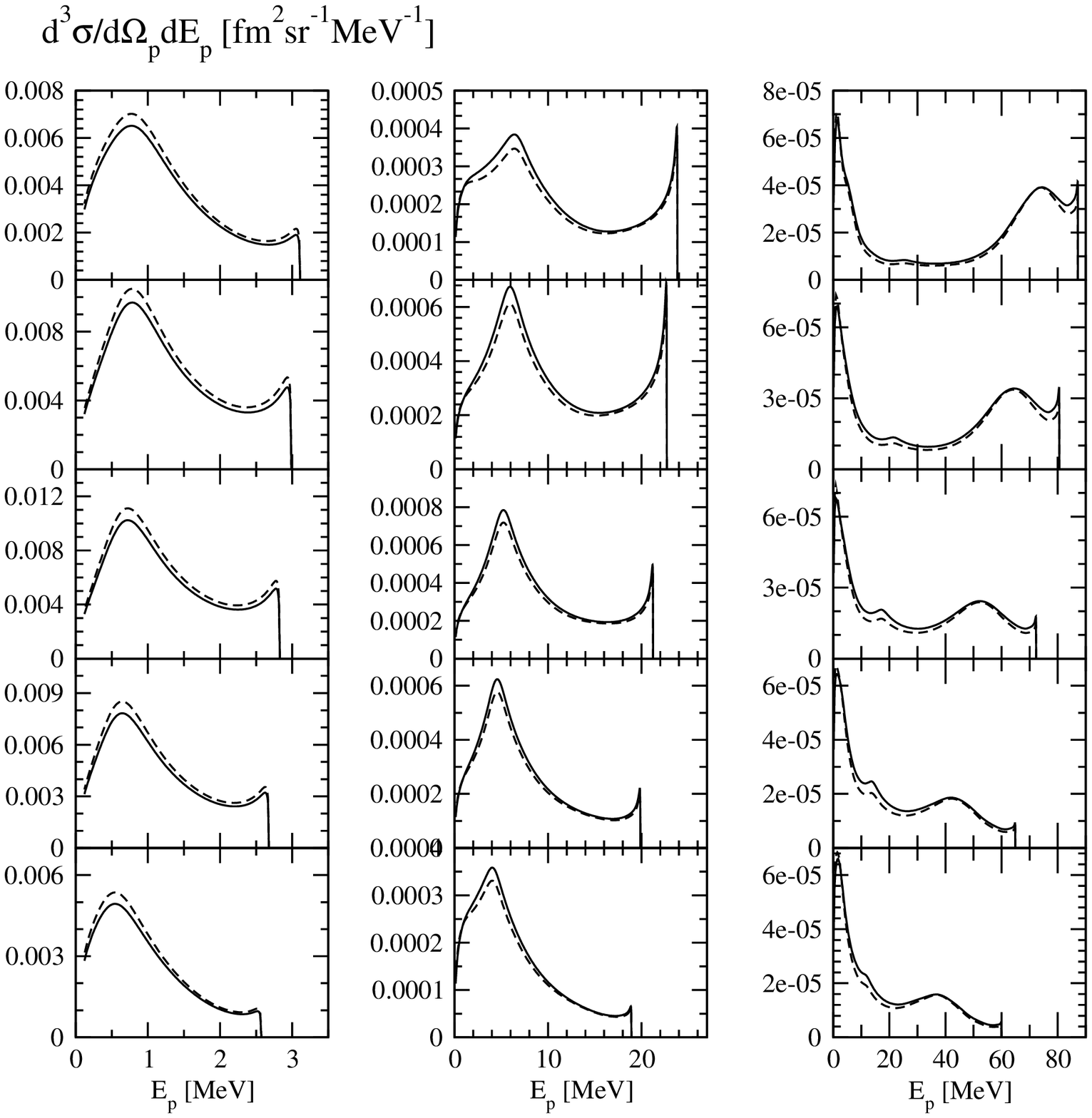}
}}
\caption[ ]
{
The differential cross section $d\sigma^3/d\Omega_p dE_p$ for $E_{\gamma}$ = 12 MeV (the first column), 40 MeV
(the second column) and 120 MeV (the third column) at different outgoing proton angles.
The first, second, third , forth and fifth row correspond to the detection angle $\theta_p$=
30$^\circ$, 60$^\circ$, 90$^\circ$, 120$^\circ$ and 150$^\circ$, respectively.
The dashed (solid) curve represents the AV18 (AV18+Urbana IX) predictions.
}
\label{fig1crossthp}
\end{figure}

\begin{figure}[h!]
\leftline{\mbox{
\epsfysize=210mm \epsffile{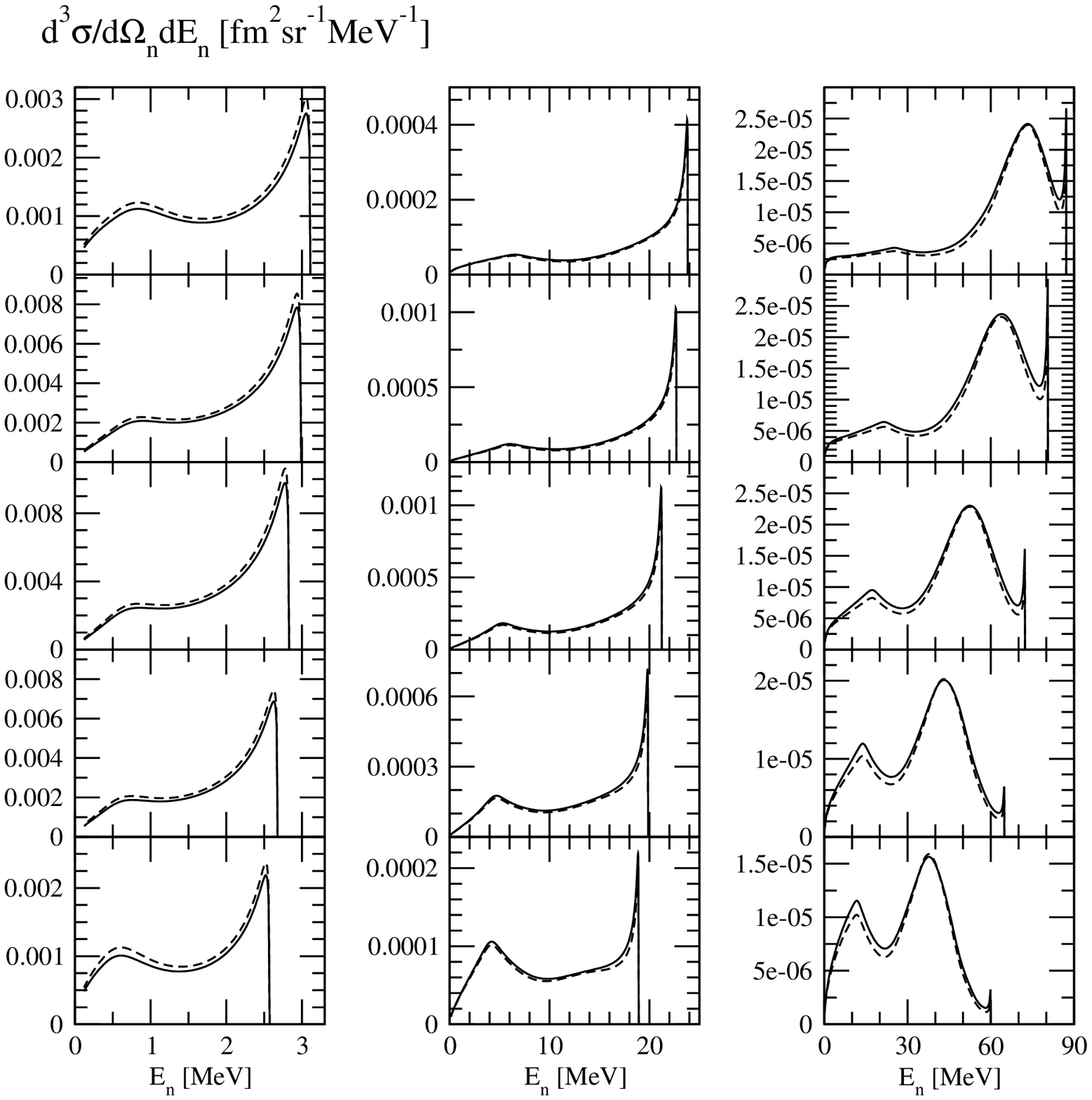}
}}
\caption[ ]
{
The same as in Fig.~\ref{fig1crossthp} but for the neutron knockout.
}
\label{fig2crossthn}
\end{figure}

Before presenting the polarization observables, for the sake of completeness, 
we would like to show the unpolarized
cross section for the $\gamma (^3{\rm He},N)NN$ reaction with the detected outgoing 
nucleon to be a proton (Fig.~\ref{fig1crossthp}) or a neutron (Fig.~\ref{fig2crossthn}).
We choose the detection polar angle $\theta$ 
to be $\theta= 30^\circ$, 60$^\circ$, 90$^\circ$, 120$^\circ$ or 150$^\circ$.
The spectra at $\theta=90^\circ$ were already presented
in ~\cite{skibinski2}. The structures seen in these spectra  originate from 
an interplay between strong final state interactions, meson exchange currents, phase space factors
and the properties of the 3N bound state wave function. For example, for the neutron spectrum 
at $E_\gamma=120$ MeV and $\theta=90^\circ$ two peaks around $E_n \approx 20$ and 70 MeV 
come from the final state interactions between two nucleons. The maximum around 50 MeV comes from the
interplay between the two-body currents, the phase space factors and the properties of the $^3$He 
bound state wave function.
As is seen in Figs.~\ref{fig1crossthp} and \ref{fig2crossthn} that structure 
depends smoothly on the angle of the outgoing nucleon
with the largest cross sections around 
$\theta=90^\circ$. The Urbana IX 3NF effects are visible at the lower and the upper energies of the incoming
photon,
and are nearly negligible at the intermediate energy.
  
\begin{figure}[h!]
\leftline{\mbox{
\epsfysize=210mm \epsffile{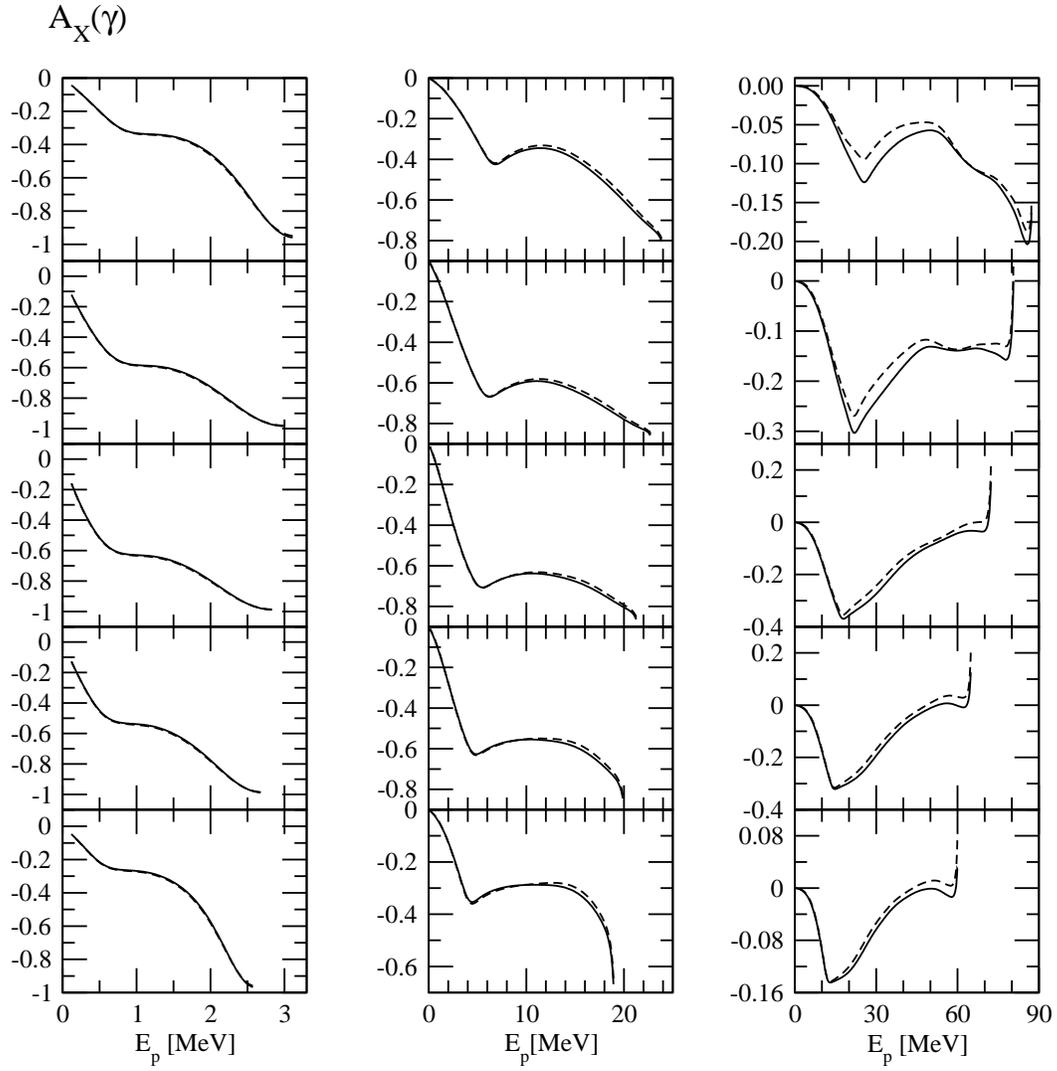}
}}
\caption[ ]
{
The photon analyzing power $A_x^{\gamma}(\theta)$ for the proton emission. 
The incoming photon energies, angles and curves are the same as in Fig~\ref{fig1crossthp}.
}
\label{fig3axgthp}
\end{figure}

\begin{figure}[h!]
\leftline{\mbox{
\epsfysize=210mm \epsffile{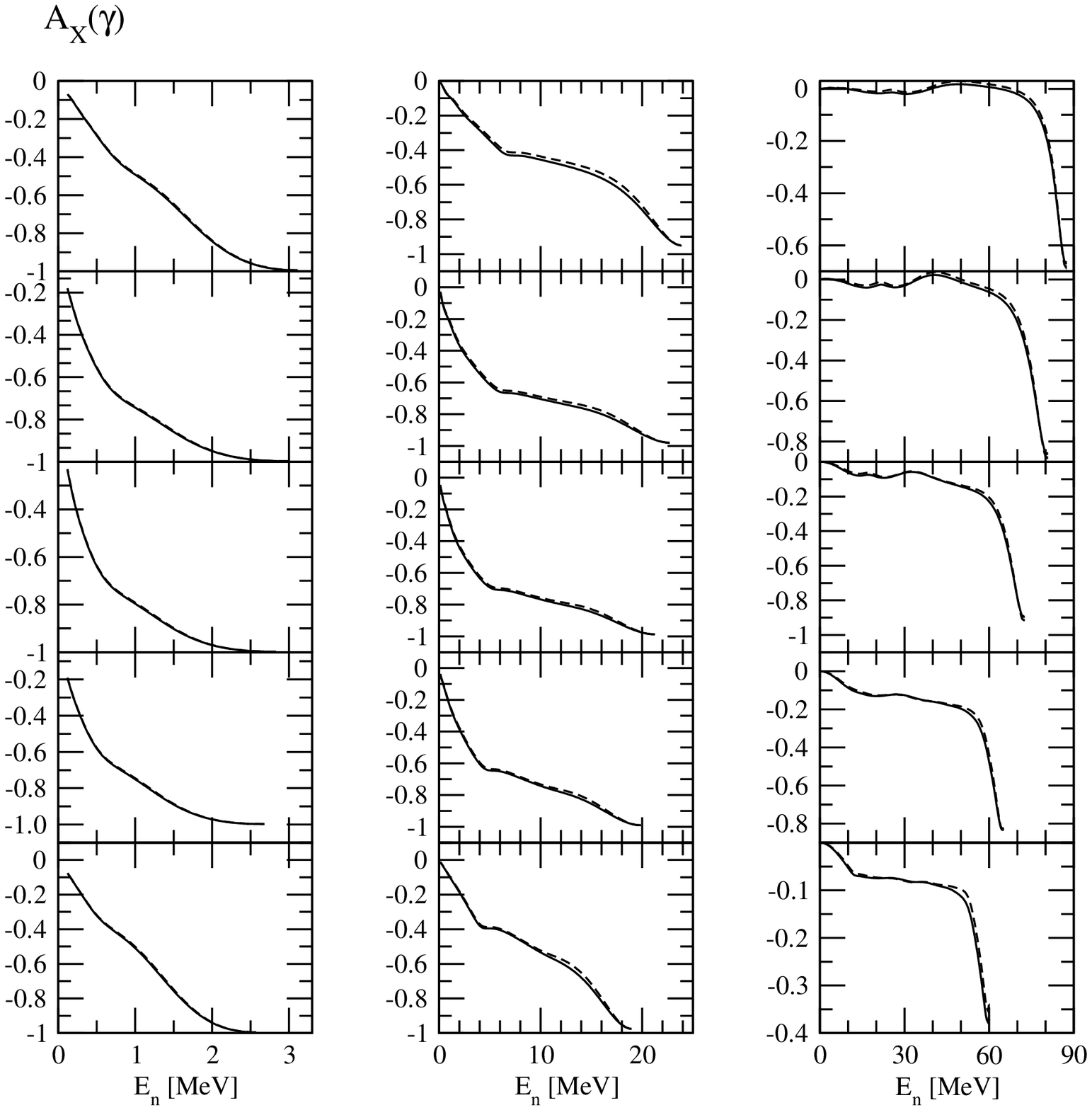}
}}
\caption[ ]
{
The same as in Fig.~\ref{fig3axgthp} but for the neutron knockout.
}
\label{fig4axgthn}
\end{figure}

The photon analyzing power $A_x^{\gamma}(\theta)$ are
shown in Figs.~\ref{fig3axgthp}-\ref{fig4axgthn}.
For photon energies $E_\gamma=12$ and 40 MeV and detecting protons this observable decreases with 
increasing proton 
energy and reaches values -1 and -0.8 at the highest proton energies, respectively.
$A_x^{\gamma}(\theta)$ is rather insensitive to the 3NF at these photon energies.
However, at $E_\gamma=120$ MeV the 3NF effects become sizable, and 
they change the photon analyzing power by up to $\approx$10\%. 
The strongest effects are visible at protons energies around 25-50 MeV and
at lower detection angles.
For the detected neutron $A_x^{\gamma}(\theta)$ reaches values -1 for $E_\gamma=12$ and 40 MeV at
the upper ends of the spectra.
At $E_\gamma=120$ MeV the value of the photon analyzing power 
is small (up to $\approx$ -0.2) except in of the region of maximal energies of the detected neutrons.
At all investigated energies the 3NF effects are negligible when the neutron is detected.

\begin{figure}[h!]
\leftline{\mbox{
\epsfysize=210mm \epsffile{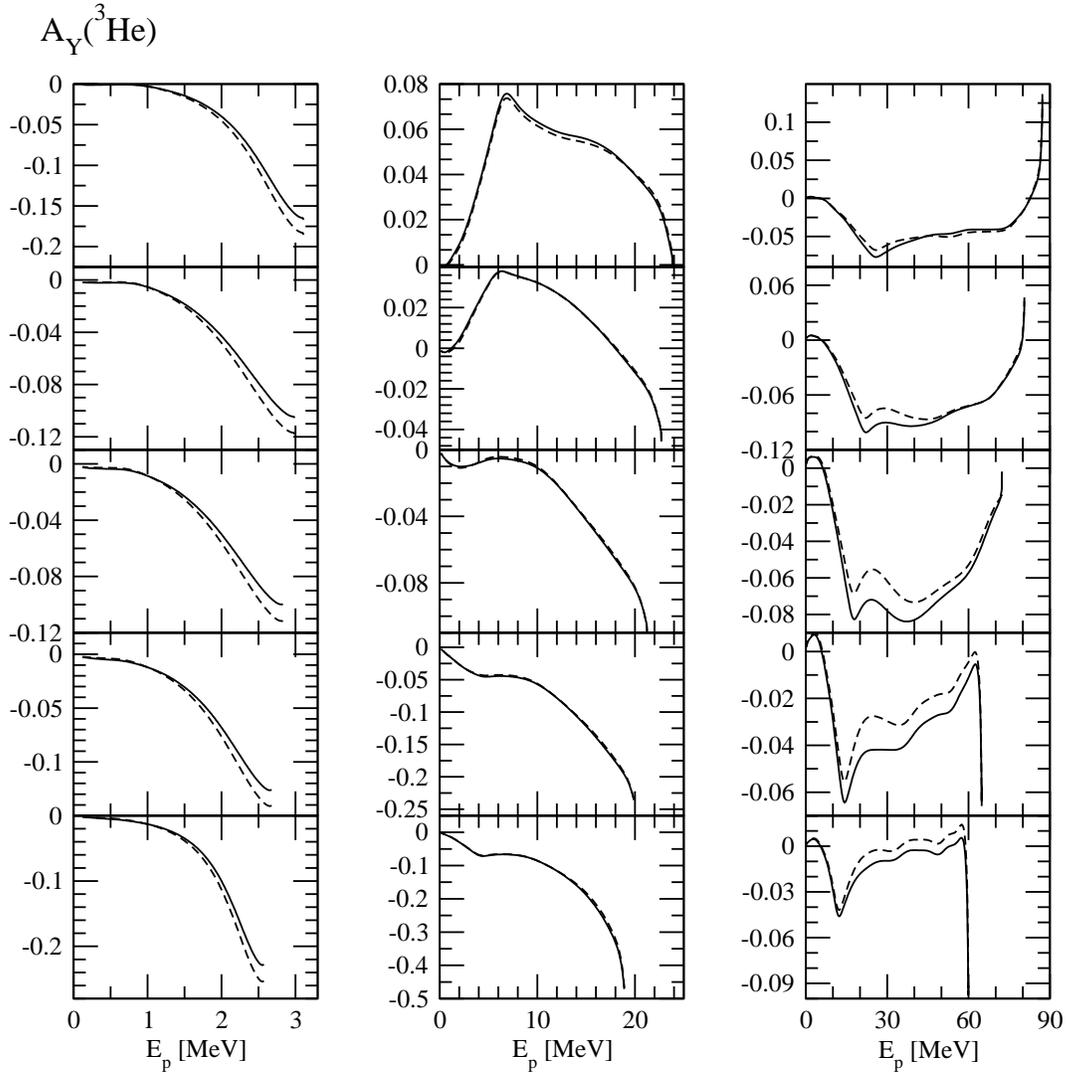}
}}
\caption[ ]
{
The $^3$He analyzing power $A_y^{^3He}(\theta)$ for the proton emission.
The photon energies, angles and the curves are as in Fig~\ref{fig1crossthp}.
}
\label{fig5ay3hethp}
\end{figure}

\begin{figure}[h!]
\leftline{\mbox{
\epsfysize=210mm \epsffile{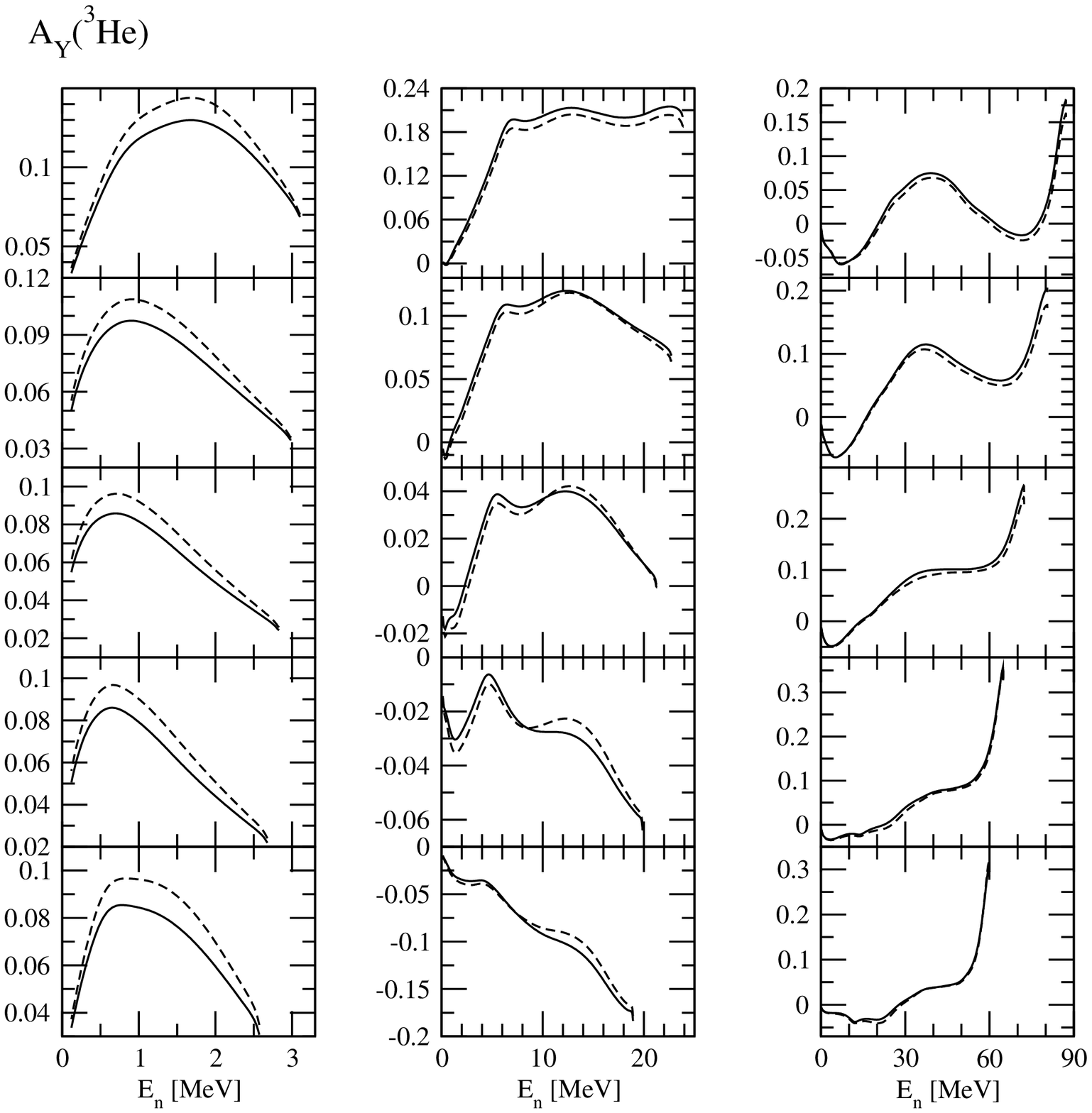}
}}
\caption[ ]
{
The same as in Fig.~\ref{fig5ay3hethp} but for the neutron knockout.
}
\label{fig6ay3hethn}
\end{figure}

Contrary to a rather small 3NF effects in the photon analyzing power, the
$^3$He analyzing power $A_y^{^3He}(\theta)$ is sensitive to the action of 3N forces (see 
Figs.~\ref{fig5ay3hethp}-\ref{fig6ay3hethn}). 
This is the case especially for the two lowest photon energies and 
the detected neutron and at $E_\gamma=12$ MeV and $E_\gamma=120$ MeV when the proton is measured.
For the detected neutron the largest 3NF effects of up to 15\% are at $E_\gamma=12$ MeV and they are
seen in the whole neutron spectrum.
In the proton case the most interesting situation is the highest photon energy $E_\gamma=120$ MeV, where
3NF effects of a magnitude above $\approx$ 20\% are seen nearly for all energies of the detected proton. 
The action of the 3NF shifts the predictions in the opposite directions for the lowest and the highest photon energy.
Unfortunately in the case of the detected proton the 3NF effects occur at 
relatively small (below 0.1) absolute values of $A_y^{^3He}(\theta)$. 
For the detected neutron 3NF effects occur also for $ A_y^{^3He}(\theta) \leq 0.12$. 
However, in this case 
3NF effects are seen even at intermediate photon energy, at
all neutron angles and in the whole energy range.
The structure of the spectrum is again due to an interplay of all dynamical components 
in the nuclear matrix elements.
The dependence of the nuclear analyzing power on the direction of the outgoing nucleon is rather smooth, but
the shape of the spectra changes significantly for different photon energies.

\begin{figure}[h!]
\leftline{\mbox{
\epsfysize=210mm \epsffile{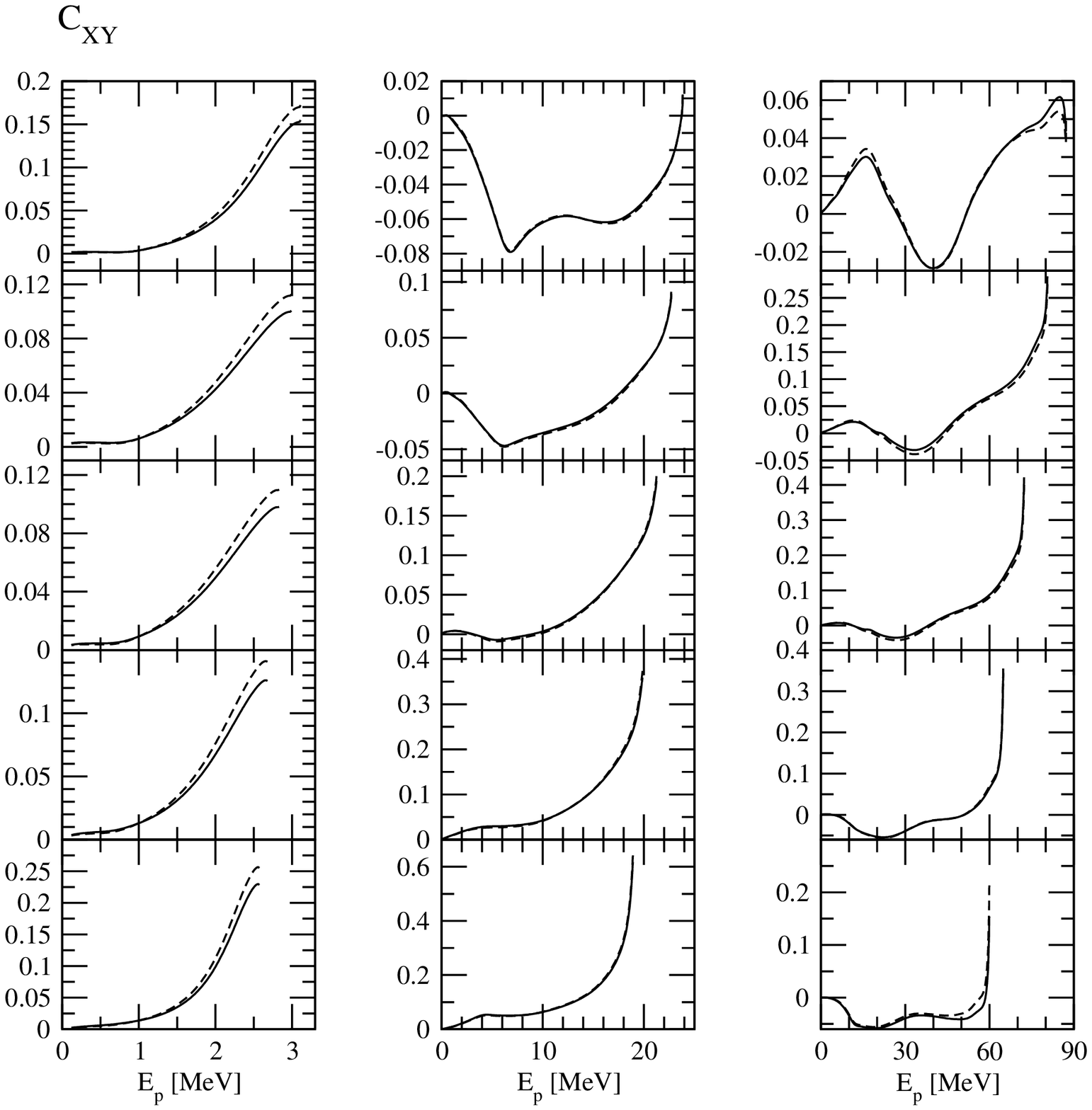}
}}
\caption[ ]
{
The spin correlation coefficients $C_{x,y}^{\gamma,^3He}(\theta)$ for the proton emission.
The energies, angles and the curves are as in Fig~\ref{fig1crossthp}.
}
\label{fig7cxythp}
\end{figure}

\begin{figure}[h!]
\leftline{\mbox{
\epsfysize=210mm \epsffile{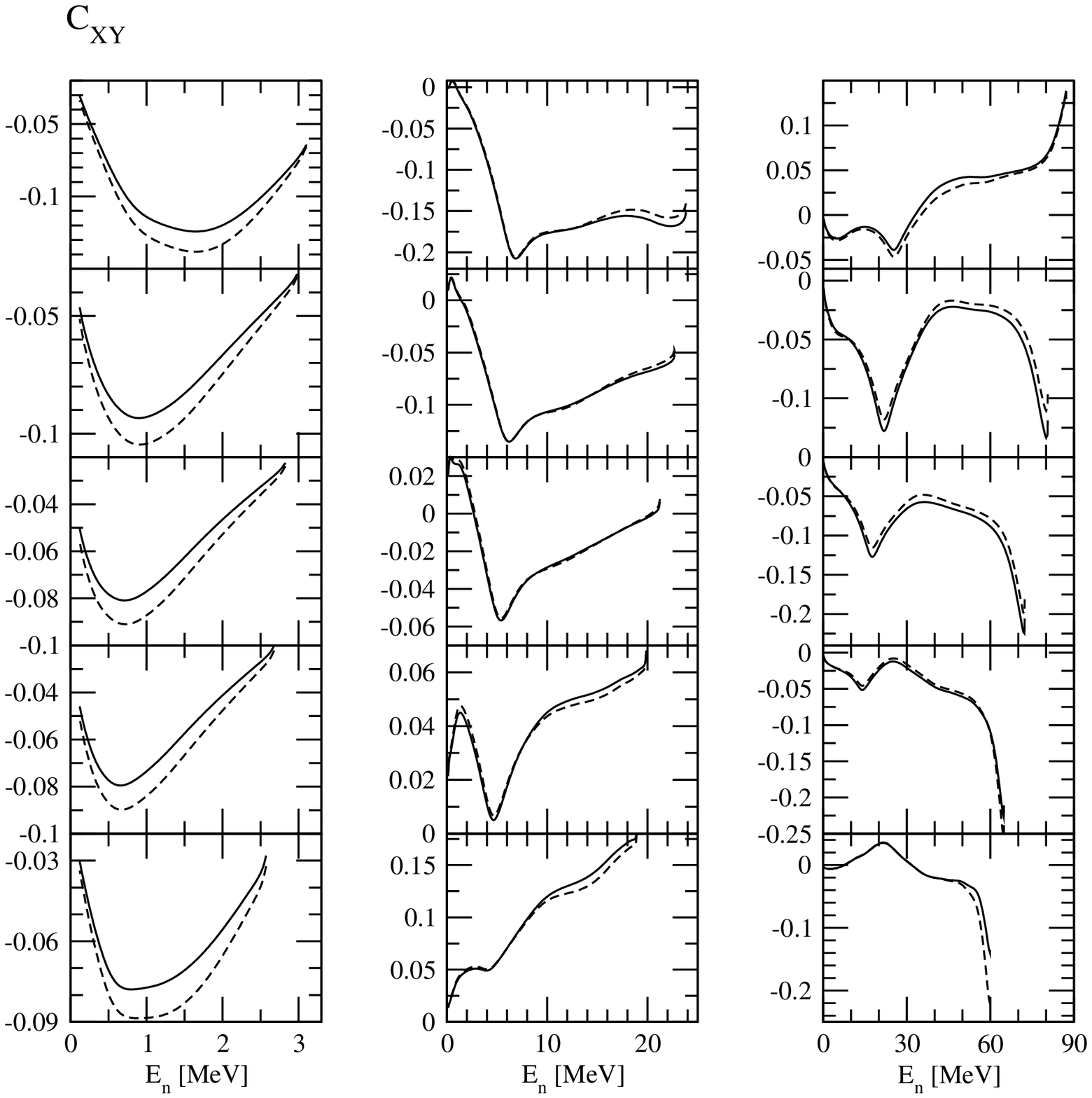}
}}
\caption[ ]
{
The same as in Fig.~\ref{fig7cxythp} but for the neutron knockout.
}
\label{fig8cxythn}
\end{figure}

The spin correlation coefficients $C_{x,y}^{\gamma,^3He}(\theta)$ are shown in Figs.~\ref{fig7cxythp}-\ref{fig8cxythn}.
In that case the largest 3NF effects ($\approx$ 15\%) occur in the whole spectrum at $E_\gamma=12$ MeV 
when the neutron is measured. 
Smaller 3NF effects are also visible at other photon energies, however, their 
magnitude depends on the detection angle.   
For the measured proton, 3NF effects are negligible at the two higher photon energies.
The absolute values of $C_{x,y}^{\gamma,^3He}(\theta)$ for the detected proton (neutron) stays below 
$\approx$0.25 ($\approx$0.1) 
at $E_\gamma=12$ MeV and $\approx$0.4 ($\approx$0.25) at the two higher photon energies.

\begin{figure}[h!]
\leftline{\mbox{
\epsfysize=210mm \epsffile{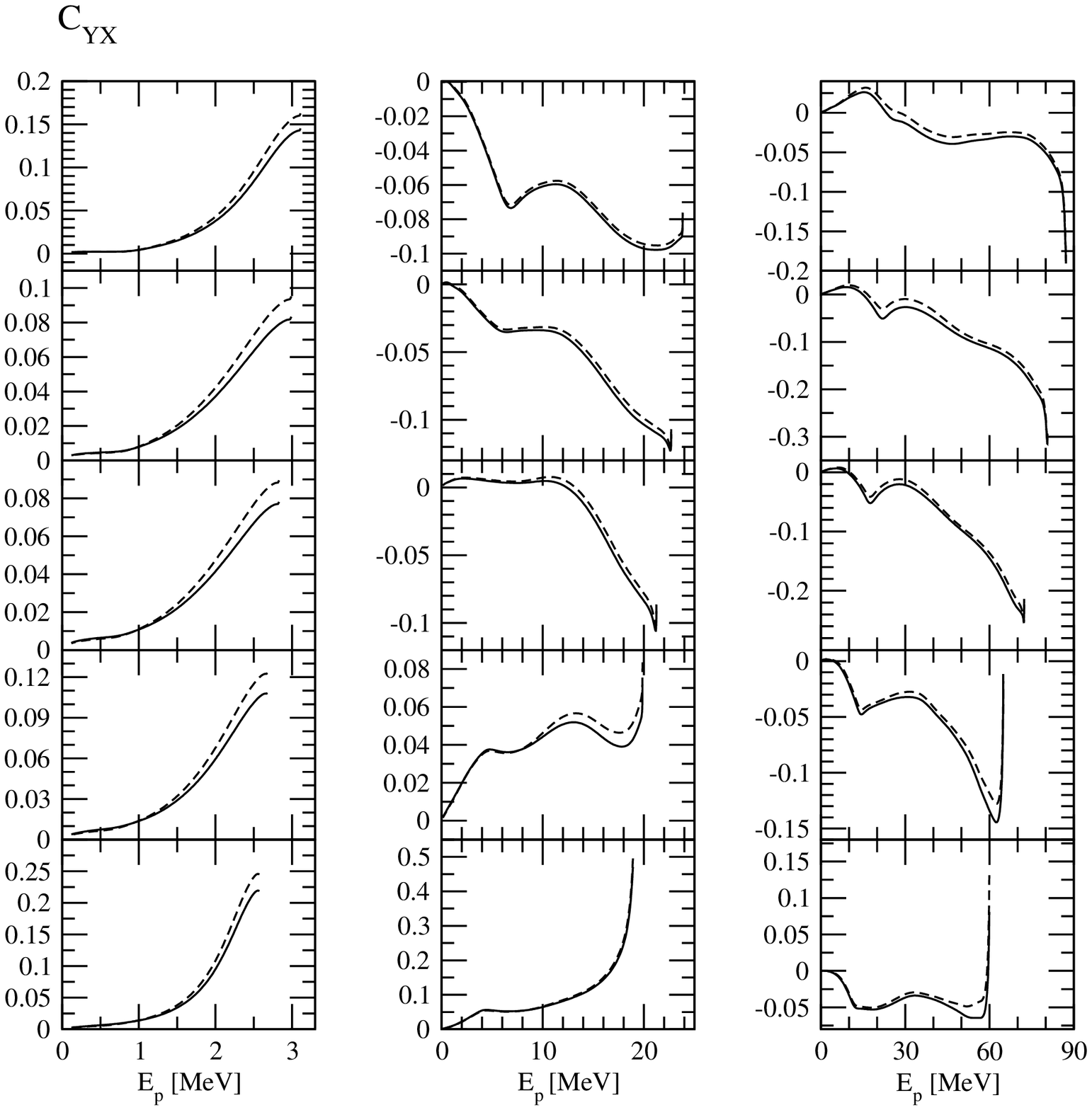}
}}
\caption[ ]
{
The spin correlation coefficients $C_{y,x}^{\gamma,^3He}(\theta)$ for the proton emission.
The photon energies, angles and the curves are as in Fig~\ref{fig1crossthp}.
}
\label{fig9cyxthp}
\end{figure}

\begin{figure}[h!]
\leftline{\mbox{
\epsfysize=210mm \epsffile{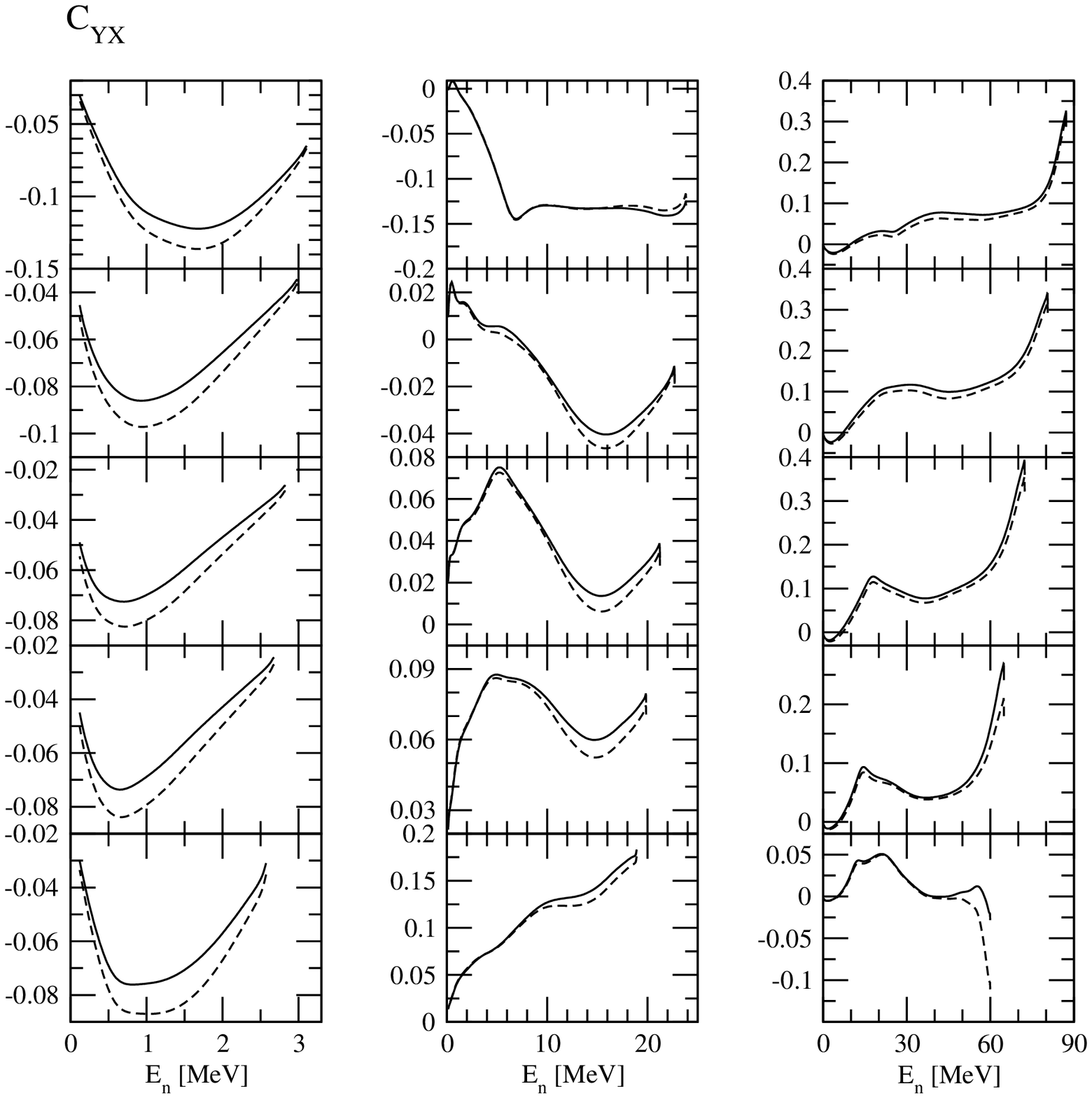}
}}
\caption[ ]
{
The same as in Fig.~\ref{fig9cyxthp} but for the neutron knockout.
}
\label{fig10cyxthn}
\end{figure}

A similar picture arises for the spin correlation coefficients $C_{y,x}^{\gamma,^3He}(\theta)$ (see 
Figs.~\ref{fig9cyxthp}-\ref{fig10cyxthn}).
Here 3NF effects are also visible at higher photon energies. For $E_\gamma=40$ MeV and the measured
neutron, 3NF effects are largest around the outgoing neutron energy $\approx$16 MeV and $\theta=60^\circ-120^\circ$.
The absolute values of $C_{y,x}^{\gamma,^3He}(\theta)$ for neutron detection are below $\approx$0.1 
for $E_\gamma=12$ and 40 MeV,
and approach up to $\approx$0.4 for $E_\gamma=120$ MeV. For the measured 
proton $C_{y,x}^{\gamma,^3He}(\theta)$ reaches 
0.25, 0.5 and 0.25 for $E_\gamma=12$, 40 and 120 MeV, respectively.
In the case of the detected proton the small 3NF effects (below 10\%) occur at all photon and 
nucleon energies and at all detection angles.

Now we would like to address the sensitivity of the spin observables to the nuclear current used.
To study this, we compare the predictions for the above spin observables 
at the detection angle $\theta=90^\circ$ for three different choices of the current operator:
the single nucleon current (SNC) only,
when the explicit two-body 
meson exchange currents are added to the SNC, and
finally when the current operator is constructed using  
the Siegert theorem~\cite{golak2000}.
The Siegert approach will also include 3N currents in the electric multipoles. We should mention, however,
that in our realization of the Siegert theorem~\cite{golak2000} we keep only single nucleon operators and do not
(yet) supplement the magnetic multipoles by the explicit $\pi$- and $\rho$- exchange currents.
Also the explicit $\pi$- and $\rho$- currents are not fully consistent with the underlying AV18 NN force,
but only with its dominant parts~\cite{Carlson98}.
For a recent investigation filling that gap see~\cite{marcucci2005}.
Despite these defects we think that the comparison of our Siegert approach
with the explicit use of the $\pi$- and $\rho$- currents will enable us to identify 
those observables,
which are especially sensitive to the choice of two and possibly three-body currents.
 
\begin{figure}[h!]
\leftline{\mbox{
\epsfysize=190mm \epsffile{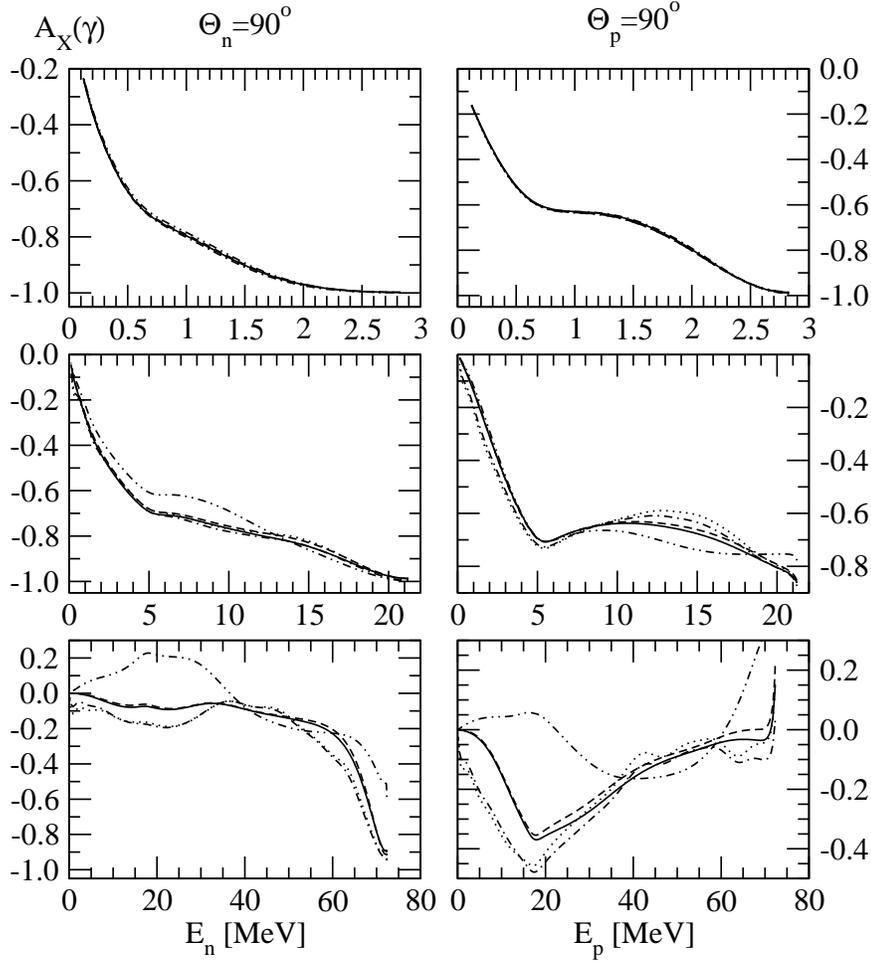}
}}
\caption[ ]
{
The photon analyzing power $A_x^{\gamma}(\theta)$ for the neutron (left column) and the proton (right column) at
$E_{\gamma}$ = 12 MeV (the first row), 40 MeV (the second row) and 120 MeV (the third row).
The nucleon detection angle is $\theta =90^\circ$. 
The double-dot-dashed line corresponds to AV18 predictions with nuclear current taken as single nucleon current only.
The dashed (solid) line corresponds to AV18 (AV18+UrbanaIX) predictions based on single nucleon current supplemented by
$\pi-$ and $\rho-$ meson exchange currents.
The dotted (dash-dotted) line corresponds to AV18 (AV18+UrbanaIX) predictions with many-body contributions to the
current taken into account via the Siegert theorem.
}
\label{fig11axg}
\end{figure}

The photon analyzing power $A_x^{\gamma}(\theta)$ is
insensitive to such a change of the nuclear current at the lowest energy (see Fig.~\ref{fig11axg}). 
At $E_\gamma=40$ MeV only a slight shift of predictions is observed under inclusion of
the meson exchange currents. The effects coming from the two models of exchange currents are insignificant
for the neutron knockout but lead to a small spread of theoretical
predictions for the proton detection.
At $E_\gamma=120$ MeV one finds  
a clear difference when the two models including exchange currents are used,
and when only the single nucleon current is taken into account. The difference 
between SNC predictions and explicit $\pi$- and $\rho$- currents (Siegert) 
results ammounts up to 140\% (180\%) at E$_n \approx$20 MeV,
and up to 650\% (880\%) at E$_p \approx$17 MeV, respectively.

\begin{figure}[h!]
\leftline{\mbox{
\epsfysize=210mm \epsffile{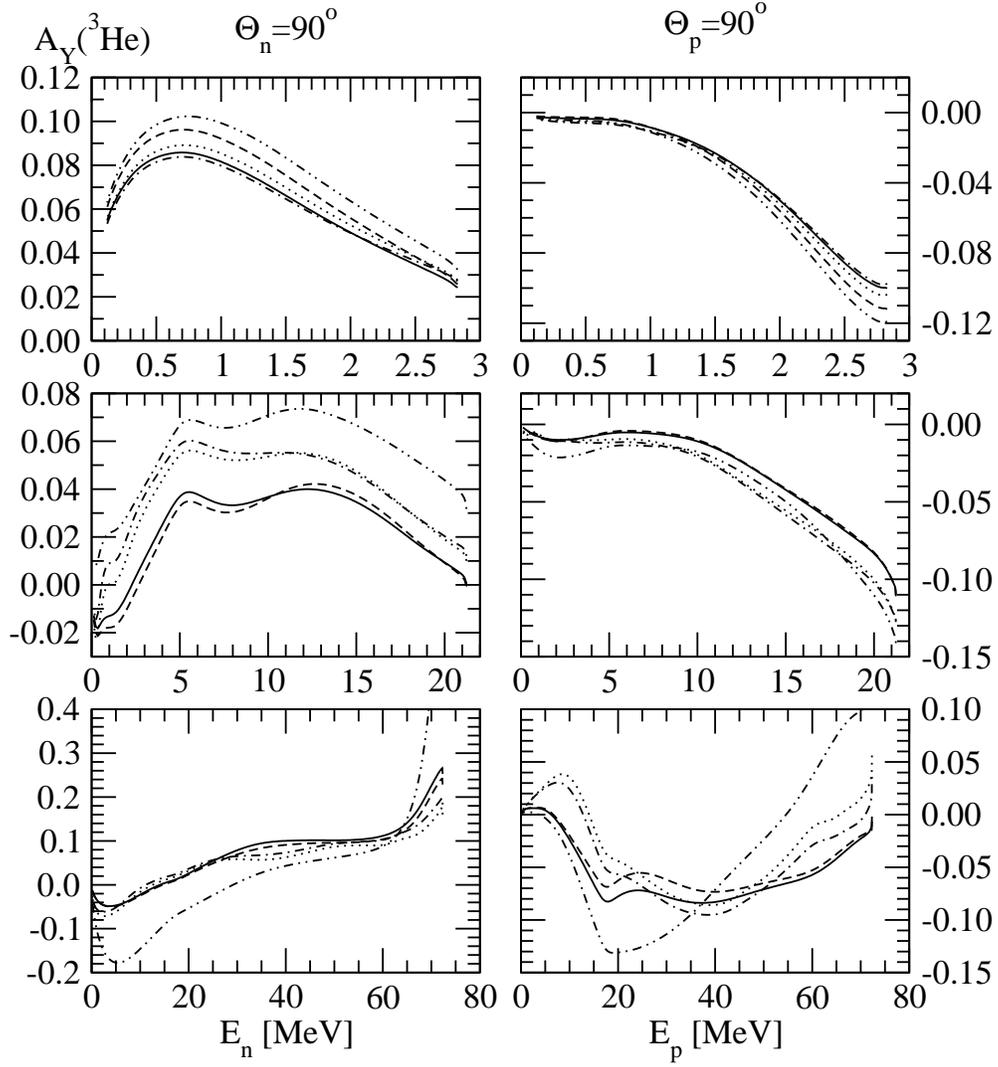}
}}
\caption[ ]
{
The same as in Fig.~\ref{fig11axg} but for the nuclear analyzing power $A_y^{^3He}(\theta)$.
}
\label{fig12ay3he}
\end{figure}

\begin{figure}[h!]
\leftline{\mbox{
\epsfysize=210mm \epsffile{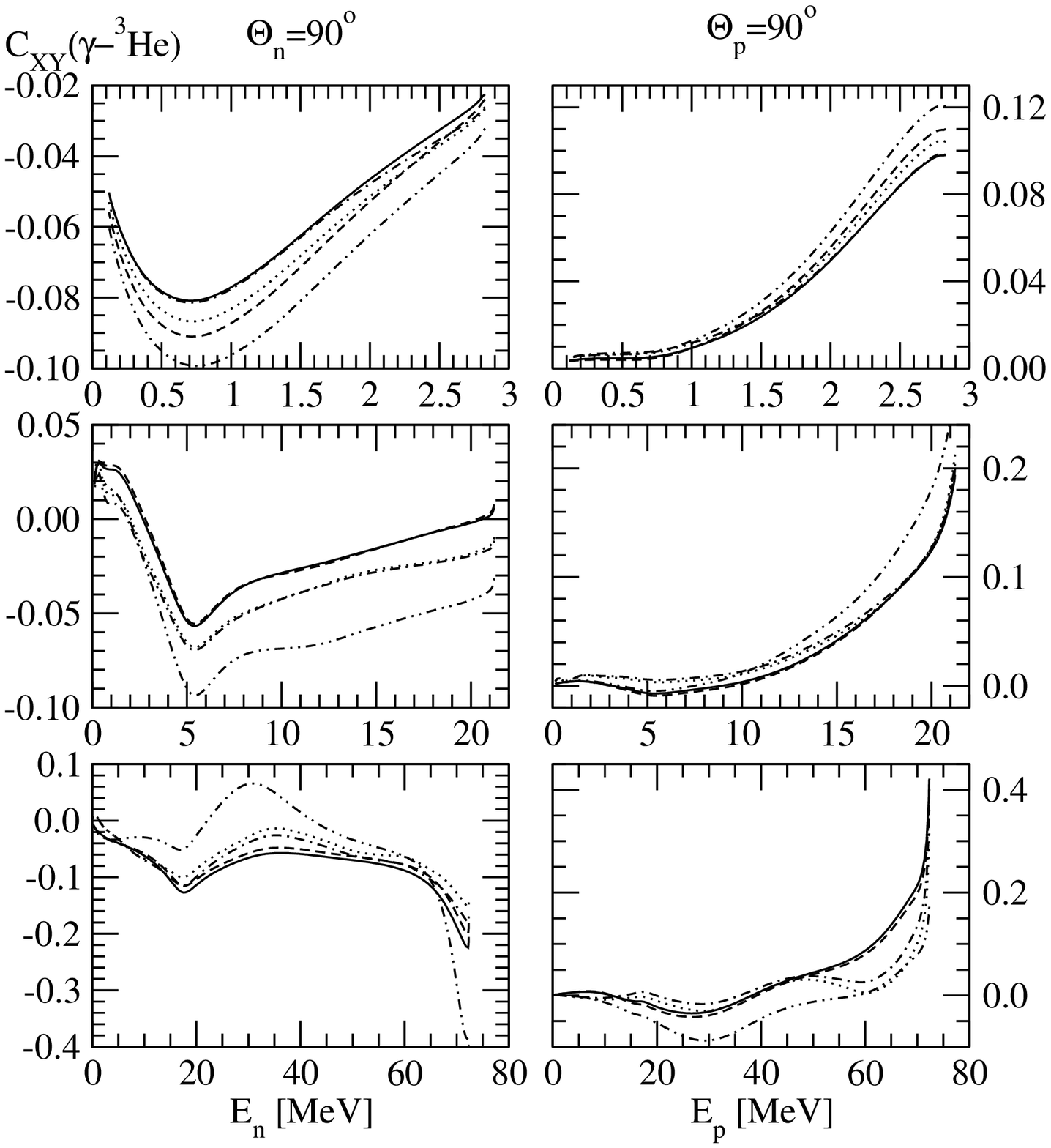}
}}
\caption[ ]
{
The same as in Fig.~\ref{fig11axg} but for the spin correlation coefficients $C_{x,y}^{\gamma,^3He}(\theta)$.
}
\label{fig13cxy}
\end{figure}

\begin{figure}[h!]
\leftline{\mbox{
\epsfysize=210mm \epsffile{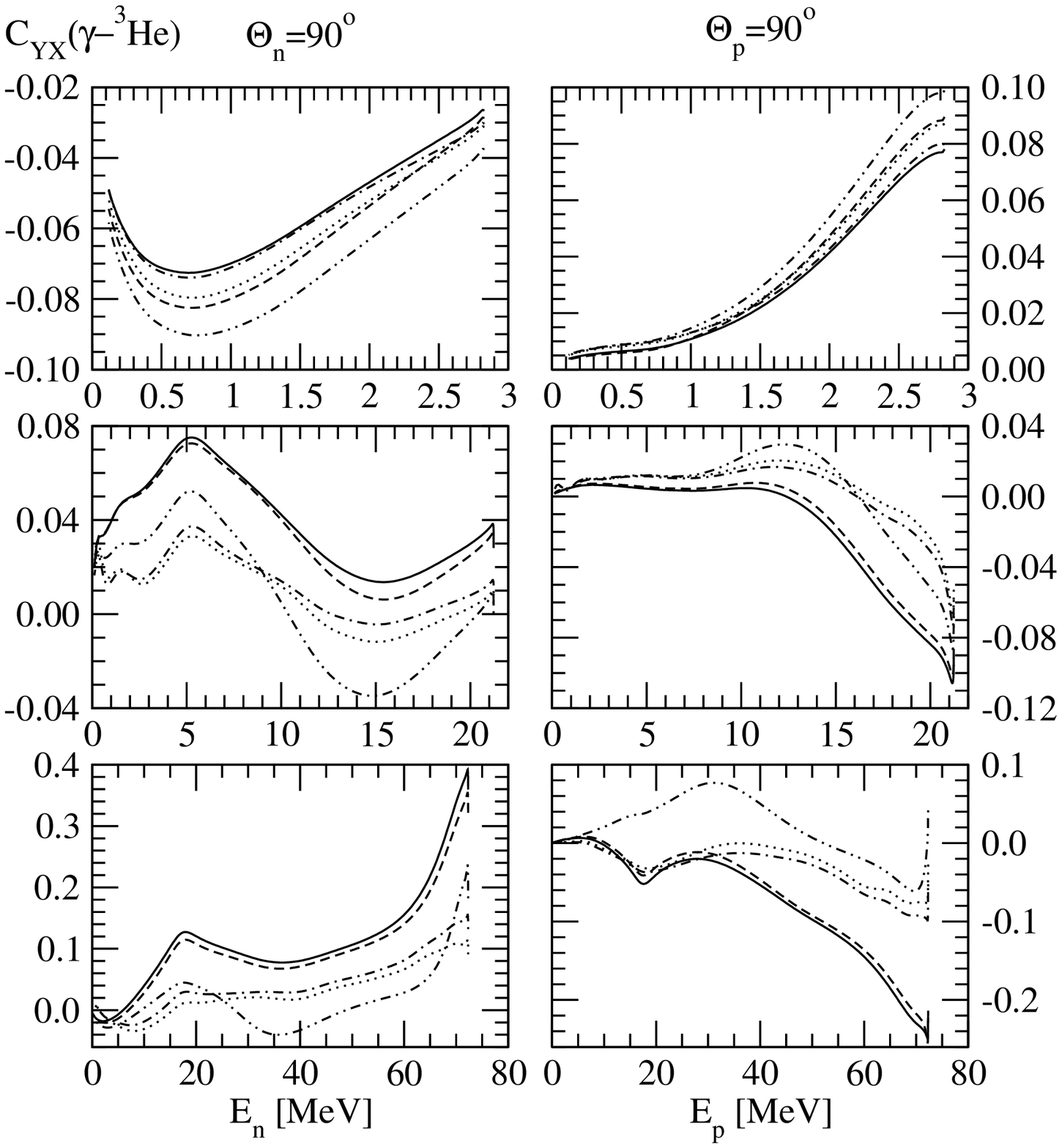}
}}
\caption[ ]
{
The same as in Fig.~\ref{fig11axg} but for the spin correlation coefficients $C_{y,x}^{\gamma,^3He}(\theta)$.
}
\label{fig14cyx}
\end{figure}

For $A_y^{^3He}(\theta)$, shown in Fig.~\ref{fig12ay3he} 
the single nucleon current predictions differ from others at all photon energies.
While for the detected neutron meson exchange currents play an 
important role at all studied energies, in the proton case they are important only at 
$E_\gamma=120$ MeV. 
The differences between Siegert and MEC are visible at 
all photon energies. At $E_\gamma=40 $ MeV they reach up to $\approx$50\% for neutron
energies around 5-10 MeV. The case of the measured proton around
$E_p \leq 15$ MeV and for $E_\gamma=120$ MeV seems to be very interesting, 
since the different nuclear currents lead to a different sign 
of $A_y^{^3He}(\theta)$ (see Fig.~\ref{fig12ay3he}). The differences are also seen for   
the spin correlation coefficients $C_{x,y}^{\gamma,^3He}(\theta)$ and $C_{y,x}^{\gamma,^3He}(\theta)$,
presented in Figs.~\ref{fig13cxy}-\ref{fig14cyx}. For $C_{x,y}^{\gamma,^3He}(\theta)$ and the measured neutron
there are clear differences, when using Siegert approach or direct $\pi-$ and $\rho-$ currents. They  
amount up to $\approx$50\% at $E_\gamma=40$ MeV. For both cases, the neutron or
proton detection, and 
$E_\gamma=12$ MeV the predictions without
3NF differ significantly, while the inclusion of the Urbana IX force leads to an agreement between both 
predictions. 
Both spin correlation coefficients are strongly influenced by the meson exchange currents.
Even at $E_\gamma=12$ MeV single nucleon current predictions differ significantly from results based 
on the nuclear current supplemented by exchange currents. The role of exchange currents
grows with the photon energy. In the case of $C_{y,x}^{\gamma,^3He}(\theta)$, $E_\gamma=40$ MeV and 
proton energies below $E_p \leq 9$ MeV, we observe different action of the
exchange currents when they are included via Siegert or by the explicit $\pi$- and $\rho-$ exchanges. 
It shows that this observable is very interesting
to study details of the nuclear current operator and deserves experimental efforts.
Thus we can state that the spin observables in the
$\vec{\gamma} (\vec{^3{\rm He}},N)NN$ reaction can provide valuable information about the nuclear 
current operator.  

\begin{figure}
\leftline{\mbox{
\epsfysize=80mm \epsffile{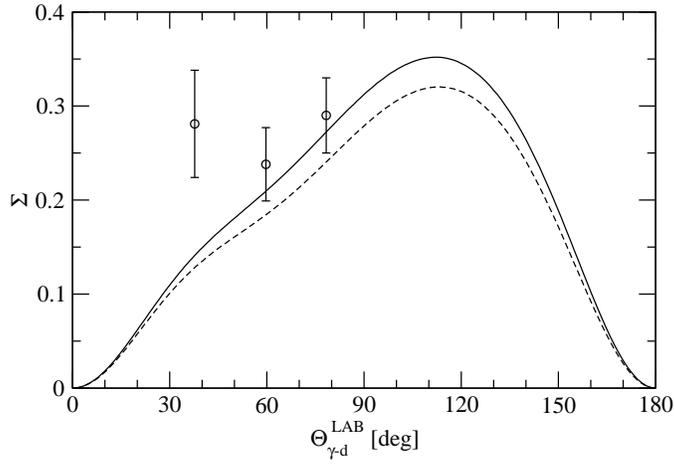} 
}}
\caption[ ]
{
The cross section asymmetry $\Sigma$ at $E_{\gamma}$ = 120 MeV.
The dashed (solid) curve represents the AV18 (AV18+Urbana IX) predictions.
Data are from~\cite{belyaev}.
}
\label{fig11ganenko120}
\end{figure}

\begin{figure}
\leftline{\mbox{
\epsfysize=80mm \epsffile{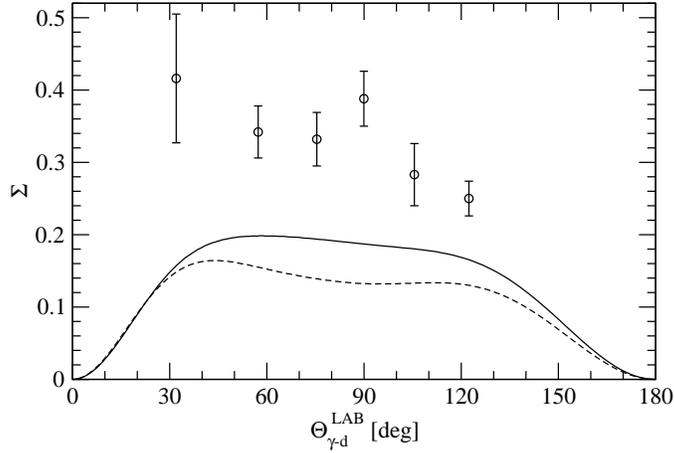}
}}
\caption[ ]
{
The cross section asymmetry at $E_{\gamma}$ = 200 MeV.
Curves as in Fig.~\ref{fig11ganenko120}
Data are from~\cite{belyaev}.
}
\label{fig12ganenko200}
\end{figure}

Finally, we address ourselves to the $\vec{\gamma} (\vec{^3{\rm He}},p)d$ process and 
compare our results with the data of Ref.~\cite{belyaev}. 
There the cross section asymmetry 
\begin{equation}
\Sigma \equiv \frac{d\sigma_\parallel - d\sigma_\perp}{d\sigma_\parallel + d\sigma_\perp},
\end{equation}
where $d\sigma_\parallel$ ($d\sigma_\perp$) is the cross section measured  
parallely (perpendicularly) to the photon polarization direction,
was investigated for linearly polarized photons with energies above 90 MeV. 
In Fig.~\ref{fig11ganenko120} we compare our predictions to the data of~\cite{belyaev}
at the photon energy $E_\gamma=120$ MeV. We see that two of the three data points are in 
good agreement with our theory. Our  prediction at the third data point is 
too low in comparison to data. The 3NF shifts the theory in the right direction into the two data points. 
Unfortunately,
most of the data points taken in~\cite{belyaev} are at photon energies above the pion 
production threshold where
our formalism is not adequate.   
Nevertheless, in Fig.~\ref{fig12ganenko200} we compare our predictions with data at $E_\gamma=200$ MeV in order
to check if our predictions at higher energies give at least a qualitative description of the data.
We see that while the shape of the theoretical predictions is similar to the shape of the data,
the absolute values of the predicted analyzing power are too small by a factor of 2.
This probably can be traced back to the missing dynamical ingredients in our theoretical framework, 
which may become  important at
such high energies. 
As was the case at $E_\gamma=120$ MeV, also at $E_\gamma=200$ MeV the 3NF improves the description of the data.
Since our calculations are much more advanced than the one used in Ref.~\cite{belyaev},
we would like to point out, that the very good description of the data presented in ~\cite{belyaev}
might be to some extent accidental.

\section{Summary}
We investigated all the nonvanishing 
spin observables
in the three-body, semiexclusive $^3$He photodisintegration process when the incoming photon and/or 
the $^3$He target nucleus
are polarized.
We found that the dependence of those spin observables on the angle of the outgoing nucleon 
is rather smooth and in most cases the shape of the energy spectra slightly changes with the incoming 
photon energy.
In the case of the $A_y^{^3He}(\theta)$ analyzing power and the spin correlation coefficients $C_{x,y}^{\gamma,^3He}(\theta)$
and $C_{y,x}^{\gamma,^3He}(\theta)$
clear effects of the 3NF are seen.  
Some of the observables (e.g. $C_{y,x}^{\gamma,^3He}(\theta)$) are sensitive to the 
details of the many-body contributions to the nuclear current operator, which we 
examplified by using the single nucleon current alone and by supplementing it either with explicit 
inclusion of $\pi$- and $\rho$- meson exchange currents
or by applying the Siegert theorem.

The presented results show that the polarization observables for $^3$He photodisintegration,
even in the relatively simple semiexclusive experiments, could provide valuable data
to test the nuclear forces and/or the reaction mechanism. 
These observables should be studied experimentally.
On the other hand, there are observables (e.g. $A_x^{\gamma}(\theta)$ at E$_\gamma$=12 MeV)
which are insensitive to the choosen current operator model and to the inclusion
of the 3N force. Such observables are natural candidates to test the most simple dynamical ingredients. 

\label{secIV}

\acknowledgements
This work was supported by
the Polish Committee for Scientific Research
under Grant No. 2P03B0825 and NATO Grant No. PST.CLG.978943.
W.G. would like to thank the Polish-German Academical Society.
The numerical calculations have been performed
on the Cray SV1 and IBM Regatta p690+ of the NIC in J\"ulich, Germany.

\newpage

\newpage

\newpage

\newpage

\newpage

\newpage

\newpage

\newpage

\newpage

\newpage

\newpage

\newpage

\newpage

\newpage

\end{document}